\documentstyle[prl,aps]{revtex}
\begin{document}
\draft

\twocolumn[\hsize\textwidth\columnwidth\hsize\csname@twocolumnfalse\endcsname                            %

\title{$\Lambda=0$ Cosmology of a Brane-like Universe}
\author{Aharon Davidson}
\address{Physics Department, Ben-Gurion University of the Negev,
Beer-Sheva 84105, Israel \\
(davidson@bgumail.bgu.ac.il)}
\maketitle

\begin{abstract}	
We examine the possibility that Friedman-Robertson-Walker
evolution is governed by an effective (rather then by the
actual) energy density.
A concrete example is provided by $\Lambda=0$ Regge-Teitelboim
cosmology, where critical cosmology only requires subcritical
matter density $(\Omega_{m}<1)$, and the age of a flat matter
dominated Universe gets enhanced by a factor of $\,\frac{9}{8}$.
Dual to the mature dilute Universe is the embryo Universe, the
evolution of both is governed by $P_{eff}=-\frac{1}{9}\rho_{eff}$.
\end{abstract}
]

\medskip
\noindent \textit{\textbf{Introduction:}}

Simple inflation\cite{Guth} models predict that the present
expansion should exhibit near flatness\cite{early}, that is
$\Omega_{0}\approx 1$.
Stretching the cosmological dark matter option to its limits
can only account for $\Omega_{m}\simeq 0.2 - 0.4$\cite{omega},
a value which roughly agrees with galactic data and virial
estimates.
It seems that the only way $\Omega_{0}=1$ can be reconciled
with observation is by the existence of some smooth (unclustered)
component.
No wonder the leading cosmological model\cite{CDM} (see\cite{exotic}
for more exotic possibilities) has revived Einstein's 'biggest
blunder', the cosmological constant $\Lambda$.
However, the naive idea of $\Omega_{m}+\Omega_{\Lambda}\approx 1$,
while numerically tenable, has created the worst fine-tuning
problem ever.
An updated cold dark matter analysis\cite{xCDM} seems to favor a
smooth $P=-0.6\rho$ background.
The alternative being a $\Lambda=0$ open Universe model\cite{open}
that manages Linde-style\cite{Linde} not to contradict the spirit
of inflation.
The debate between the various models is usually taken to the
$H_{0}-\Omega_{0}$ plane, where $H_{0}=100 h \,\mathrm{kms^{-1}
Mpc^{-1}}$ is the present Hubble rate.

The age $t_{0}$ of the Universe is another key factor in the game.
A paradoxical situation where the expanding age of the Universe is
smaller than the age of some of its constituents is unacceptable.
The longer the $t_{0}$, the smaller the $H_{0}$ needed to explain
it, but the latter seems to be confined to the experimental window
$0.5\leq  h\leq 0.7$\cite{hwin}.
On the other hand, the age of the oldest globular clusters is
now estimated\cite{12Gyr} to be $\sim 12$ Gyr.
Appealing to the lower bounds, this may be at the edge of
consistency\cite{Krauss}; a $10\%$ enhancement mechanism for
$t_{0}h$ is however quite welcome.
As nicely stated by Primack\cite{Primack}, the
age$\leftrightarrow$expansion-rate conflict, if exists, goes to
the heart of general relativity (GR).
Abandoning the GR trail, and conformal gravity\cite{conformal}
is a remarkable example in this respect, the rules of the game
are changed.

In this paper, we examine the possibility that the
Friedman-Robertson-Walker (FRW) evolution is governed by
an effective, rather then by the actual, energy density.
We demonstrate, in the context of a cosmological constant free
($\Lambda=0$) Regge-Teitelboim (RT) gravity\cite{RT}, how to live
with $\Omega_{matter}<1$ without sacrificing the inflation inspired
near-critical evolution.

\medskip
\noindent \textit{\textbf{Regge-Teitelboim type cosmology:}}

RT gravity has been proposed with the motivation that the first
principles which govern the evolution of the entire Universe cannot
be too different from those which determine the world-line (world-sheet)
behavior of a point particle (string).
The idea was criticized\cite{Deser} in the past, with the focus
on the gauge dependence, but has been reconsidered by several
authors\cite{RTmodels}.
The generalized RT-type action looks formally conventional
\begin{equation}
	S=\int_{}^{}(-\frac{1}{16\pi G}
	{\mathcal{R}}+{\mathcal{L}}_{m})\sqrt{-g}d^{n}x ~.
	\label{Lagrangian}
\end{equation} 
Note that, in the original RT-version, the dynamics was supposed
to enter via some first class purely geometric constraints, without
appealing to ${\mathcal{L}}_{m}$. 
At any rate, the Einstein equations get drastically modified once
$g_{\mu\nu}(x)$ is not regarded a canonical field.
This role is played here by the higher dimensional embedding
functions $y^{M}(x)$, such that $g_{\mu\nu}(x)=\eta_{MN}y^{M}_{,
\mu}y^{N}_{,\nu}$.
It is thus important to emphasize that \textit{all other equations
of motion remain absolutely intact}.
Performing now the variation with respect to $y^{M}(x)$, one faces
a set of conservation laws $[({\mathcal{R}}^{\mu\nu}-\frac{1}{2}
g^{\mu\nu}{\mathcal{R}}-8\pi GT^{\mu\nu})y^{M}_{;\mu}]_{;\nu}=0$.
The Bianchi identity combined with the embedding identity $\eta_{MN}
y^{M}_{;\lambda}y^{N}_{;\mu\nu}\equiv 0$ then imply that $T^{\mu\nu}$
\textit{is conserved}!
This is a crucial result, especially when Einstein equations
are not at our disposal.
The RT field equations, a weaker system (only six independent 
equations) in comparison with Einstein equations, then take the
compact form
\begin{equation}
	\left({\mathcal{R}}^{\mu\nu}-\frac{1}{2}g^{\mu\nu}
	{\mathcal{R}}-8\pi GT^{\mu\nu}\right)y^{M}_{;\mu\nu}=0
	\label{RTeq}
\end{equation}
Clearly, every solution of Einstein equations is automatically a
solution of the corresponding RT equations.

The induced line element is assumed to take the idealized FRW form
\begin{equation}
	ds^{2}_{FRW}=-dt^{2}+\frac{R^{2}(t)}
	{(1+\frac{1}{4}kr^{2})^{2}}\delta_{ij}dx^{i}dx^{j} ~.
	\label{FRW}
\end{equation}
Following the isometric embedding theorems\cite{theorems}, at most
$N=\frac{1}{2}n(n+1)$ background flat dimensions are required to
\textit{locally} embed a general $n$-metric.
This number can be reduced, however, if the $n$-manifold admits
some Killing vector fields.
In particular, for $n=4$, if the Universe happens to be homogeneous
and isotropic, a total number of $N=5$ flat dimensions suffices.
For the (say) $k>0$ case, to be a bit more specific but without 
losing generality, the embedding functions\cite{embedding} are
given by
\begin{eqnarray}
	y^{0}(x) & = & \int_{}^{t}
	\sqrt{1+\frac{\dot{R}^{2}}{k}}dt' ~,
	\nonumber \\
	y^{i}(x) & = & \frac{Rx^{i}}{(1+\frac{1}{4}kr^{2})} ~,
	\label{embedd} \\
	y^{4}(x) & = & \frac{R}{\sqrt{k}}
	\left(\frac{1-\frac{1}{4}kr^{2}}{1+\frac{1}{4}kr^{2}}\right) ~.
	\nonumber
\end{eqnarray}
When the algebraic dust settles down, it appears that all
five RT equations are essentially the same in this case.
Assuming a perfect fluid energy-momentum tensor, that is
$T^{\mu}_{\nu}=Diag(-\rho,P,P,P)$, they read
\begin{equation}
	\left(\rho-9\frac{\dot{R}^{2}+k}{8\pi GR^{2}}\right)
	\frac{\ddot{R}R}{\dot{R}^{2}+k}=
	3\left(P+\frac{\dot{R}^{2}+k}{8\pi GR^{2}}\right) ~.
	\label{RTrhoP}
\end{equation}

Eq.(\ref{RTrhoP}) is integrable; the integrability condition is
nothing but the familiar energy conservation law
\begin{equation}
	\dot{\rho}+3(\rho+P)\frac{\dot{R}}{R}=0 ~.
	\label{energy}
\end{equation}
Substituting eq.(\ref{energy}), we integrate eq.(\ref{RTrhoP}) into
$f(\rho,R,\dot{R})=\rho f_{1}(R,\dot{R})+f_{2}(R,\dot{R})=const$,
and after some algebra arrive at our master equation
\begin{equation}
	8\pi G\rho R^{3}(\dot{R}^{2}+k)^{1/2}-
	3R(\dot{R}^{2}+k)^{3/2}=-\frac{1}{9}\mu ~.
	\label{master}
\end{equation}
$\mu$ is the constant of integration which parameterizes the deviation
from Einstein limit $(\mu_{E}= 0)$.

Eq.(\ref{master}) can be casted in its final form
\begin{equation}
	\dot{R}^{2}+k=
	\frac{8\pi G}{3}\xi\rho R^{2} ~,
	\label{xiFRW}
\end{equation}
where $\xi(R)$ is a solution of the cubic equation
\begin{equation}
	\xi(\xi-1)^{2}=
	\frac{\mu^{2}}{27(8\pi G\rho)^{3}R^{8}} ~.
	\label{xi}
\end{equation}
The FRW evolution has been recovered; it is governed, however,
by the effective energy density $\rho_{eff}\equiv \xi\rho$,
rather than by the actual energy density $\rho$.
It is important to note that the two independent branches
$\xi>1$ and $0<\xi<1$ are separated by the troublesome
$\rho^{3}R^{8}\rightarrow\infty$.

\medskip
\noindent \textit{\textbf{'Missing' mass:}}

Now, suppose two physicists, one of which (E) is equipped with
the standard FRW cosmology, while the other (RT) is exposed
to the modified FRW theory, attempt to calculate the $\Omega$
parameter.
While agreeing on what is actually meant by the energy density
$\rho$ and the Hubble constant $H$, they derive different formulae
for the critical density $\rho_{c}$, and thus are in dispute with
regard to the fate of the Universe:

\noindent (i) For the conservative E-physicist,
critical evolution means
\begin{equation}
	\rho^{E}_{c}=\frac{3H^{2}}{8\pi G} ~.
	\label{rhoEc}
\end{equation}
On the theoretical side, being exposed to the idea of inflation,
he may naively expect $\Omega^{E}\approx 1$.
Unfortunately, as briefly explained earlier, his analysis reveals
a 'missing' mass puzzle $\Omega^{E}\equiv\rho/\rho^{E}_{c}<1$.

\noindent (ii) The challenger RT-physicist, on the other hand, argues
that it is ${\displaystyle \rho^{eff}_{c}\equiv\xi\rho^{RT}_{c}=\frac
{3H^{2}}{8\pi G}}$ which defines criticality.
As far as he is concerned,
\begin{equation}
	\rho^{RT}_{c}=\frac{3H^{2}}{8\pi\xi G}
	=\frac{1}{\xi}\rho^{E}_{c} ~.
	\label{rhoRTc}
\end{equation}
On the same experimental and theoretical grounds, he may favor
the elegance of
\begin{equation}
	\Omega^{RT} \equiv
	\Omega_{m} + \Delta\Omega \approx 1
	\hspace{8pt}\Rightarrow \hspace{8pt}
	\Omega^{E} \equiv \Omega_{m}=
	\frac{1}{\xi}\neq 1 ~,
	\label{nomissmass}
\end{equation}
where the apparently missing (or excessive) mass is interpreted
as a RT effect.
We will soon prove that $\Delta\Omega$ corresponds to some
effective $P_{eff}=-\frac{1}{9}\rho_{eff}$ background, analogous
to the familiar $P=-\rho$ background described by $\Omega_{\Lambda}$.
It worth emphasizing that
(i) Being a smooth (unclustered) component, it cannot serve as a
core for lensing measurements, and
(ii) Having no direct couplings to photons, it cannot serve as a
source for optical mass-to-light measurements.

A main concern of the RT-physicist is why \textit{must}
the Universe be associated with the $\xi>1$ missing-mass branch,
rather than with the $0<\xi<1$ excessive-mass branch?
To answer this question, consider first the evolution of an
apparently empty ($\rho\rightarrow 0$ fast enough) spacetime.
From eq.(\ref{xi}) we learn that $\xi_{\mathrm{empty}}
\rightarrow\infty$ in such a way that $\rho_{eff}=\xi\rho$
is finite.
The 'missing' mass effect is maximal in this case.
A necessary condition for having a dilute Universe is
\begin{equation}
	\xi \simeq 
	\frac{\mu^{2/3}}{24\pi G \rho R^{8/3}}\gg 1 ~.
	\label{dilute}
\end{equation}
The sufficient condition being: $\rho\rightarrow 0$ faster than
$1/R^{8/3}$.
Such a condition is automatically satisfied by conventional matter.
Appreciating the fact that $\rho_{eff}\sim 1/R^{8/3}$ in this case,
the RT evolution of a very dilute Universe becomes \textit{universal}.
It is governed by the effective equation of state
\begin{equation}
	P_{eff}=-\frac{1}{9}\rho_{eff} ~,
	\label{empty}
\end{equation}
where the $-\frac{1}{9}$ factor is in fact $\frac{1}{3}\left(
\frac{8}{3}\right)-1$.
Using the anthropic approach, we can now answer the above question.
The Universe \textit{must} evolve along the $\xi>1$ branch if it
is to enter the dilute stage.

The asymptotic behavior of $\xi(R)$, unlike $\rho_{eff}(R)$, 
heavily depends on the actual equation of state.
Given the prototype equation of state $P=(\gamma-1)\rho$,
for which $\rho(R)\sim 1/R^{3\gamma}$, one derives
$\xi(R)=\rho_{eff}/\rho \sim R^{3\gamma -8/3}$
in the dilute Universe approximation.
If $\gamma >\frac{8}{9}$, and the Universe expands along the
$\xi (R)>1$ trail, $\xi (R)$ becomes a monotonically increasing
function of $R$.
Given the matter-dominated nature of the present Universe,
we expect the 'missing' mass problem to get moderately worse
according to $\Omega^{E}\sim 1/R^{1/3}$ as the cosmic time
ticks on.

\medskip
\noindent \textit{\textbf{The expanding age of the Universe:}}

We carry out a prototype calculation under two standard assumptions:
(i) The present Universe is nearly flat, meaning a negligible
curvature term $k \ll H^{2}_{0}R^{2}_{0}$, and
(ii) The present Universe is matter dominated, so that
$\rho R^{3}=\rho_{0} R^{3}_{0}$.
RT-modified $\Lambda CDM$ and/or open Universe cosmology are to
be discussed elsewhere.

\noindent Applying the second assumption to eq.(\ref{xi}),
one derives
\begin{equation}
	\frac{\xi(\xi-1)^{2}}{\xi_{0}(\xi_{0}-1)^{2}}=
	\frac{R}{R_{0}} ~,
\end{equation}
leading to
\begin{equation}
	\dot{R}=R_{0}H_{0}\frac{\xi_{0}-1}{\xi-1} ~.
\end{equation}
With this in hand, we perform the integration
\begin{equation}
	t_{0} =  \int_{0}^{R_{0}}\frac{dR}{\dot{R}}=
	\int_{1}^{\xi_{0}}
	\frac{(\xi-1)^{2}(3\xi-1)}{H_{0}\xi_{0}(\xi_{0}-1)^{3}}
	d\xi ~,
\end{equation}
to obtain the expanding age
\begin{equation}
	t^{RT}_{0}=
	\frac{3}{4H_{0}}\left(1-\frac{1}{9}\Omega_{m}\right) ~,
	\label{age}
\end{equation}
where $\Omega_{m}\equiv 1/\xi_{0}$.
At the $\Omega_{m}\rightarrow 1$ limit, we do recover the
conventional result $t^{E}_{0}=\frac{2}{3}H^{-1}_{0}$.
At the $\Omega_{m}\rightarrow 0$ limit, where the 'missing'
mass effect is maximal, we approach the RT-enhanced expanding
age $t^{RT}_{0} \rightarrow \frac{3}{4}H^{-1}_{0}$.
To stay practical, note that a reasonable $\Omega_{m}\simeq 0.3$
gives rise to $t_{0}\simeq 0.73 H^{-1}_{0}$, very close to the
upper bound.
Consequently, $t_{0}\simeq 12 \,\mathrm{Gyr}$ corresponds to
$h\simeq 0.62$ (to be contrasted with $h\simeq 0.56$), notably
in the center of the experimental window\cite{hwin}, .

To further characterize the $\Lambda=0$ RT cosmology, we calculate the
so-called deceleration parameter $q \equiv -\ddot{R}R/\dot{R}^{2}$.
Starting from eq.(\ref{RTrhoP}), we substitute the matter domination
assumption $P=0$ and neglect the curvature terms to obtain
\begin{equation}
	q^{RT}_{0}=\frac{1}{(3-\Omega_{m})} ~.
\end{equation}
As expected, the upper bound $q_{0}=\frac{1}{2}$ is recovered for 
$\Omega_{m}\rightarrow 1$, but for a subcritical $\Omega_{m}$, we
approach the lower bound of $q_{0}=\frac{1}{3}$.

Our attempt to enhance the age of the universe by a factor of 
$\frac{9}{8}$, and subsequently reducing the deceleration parameter 
by $\frac{2}{3}$, may still fall short on realistic grounds.
The latter, recalling the supernova Hubble plot, may prefer the
$\Lambda \neq 0$ cosmology.
In which case, dark matter may turn out to be nothing but a 
Regge-Teitelboim artifact.

\medskip
\noindent \textit{\textbf{A bonus for a baby Universe:}}

Assuming that our dilute nearly-flat Universe is critically
expanding towards
$(R \rightarrow\infty ,\,\rho^{3}R^{8}\rightarrow 0)$, we would
like to trace its evolution backwards in time.
Of particular interest for us is of course the $R\rightarrow 0$
limit (if accessible).
Given $\xi>1$, we have only two $t\rightarrow0$ scenarios to
discuss:

\noindent $\bullet$ If the very early Universe is dominated by
'conventional' matter $\rho\sim R^{-(8/3+\ldots)}$, the
Big-Bang is characterized by $(R \rightarrow 0 ,\,\rho^{3}R^{8}
\rightarrow\infty)$.
In which case, the evolution trail starts at $\xi=1$ and
aims towards $\xi\rightarrow\infty$.
The initial evolution is then very sensitive to the details
of the actual equation of state.

\noindent $\bullet$ If the very early Universe is dominated by
'inflationary' matter $\rho\sim R^{-(8/3-\ldots)}$, the
Big-Bang is characterized by $(R \rightarrow 0 ,\, \rho^{3}R^{8}
\rightarrow 0)$.
In which case, the evolution trail starts at $\xi\rightarrow\infty $,
and after a transition point at some $\xi_{{\mathrm{min}}}>1$,
grows again towards $\xi\rightarrow\infty $.

\medskip
The second alternative is clearly the favorite one.
First, it re-establishes the mandatory link with the inflationary 
era, the ingredient invoked to induce critical evolution in the
first place.
Second, reflecting a \textit{duality} between the very early and
the very late Universes, it predicts a detailed pre-inflation era.
Amazingly, the evolution of the baby Universe and the
evolution of the mature (dilute) Universe are essentially
\textit{universal}, governed by the same effective equation of state
$P_{eff}\simeq-\frac{1}{9}\rho_{eff}$.
In an attempt to decode the 'missing' mass puzzle, we have
serendipitously probed the Big-Bang.

To see how the pre-inflation to inflation transition works, and
to appreciate the fact that both epochs are governed by the one and
the same actual equation of state $P=-\rho$ (but differ drastically
by their effective equation of state), we analyze the exact analytic
solution\cite{DP} in a positive cosmological constant background.
We calculate the quantity $V(R)\equiv-\frac{1}{3}8\pi G\xi\rho
R^{2}$, which plays the role of the potential energy in the
analogous mechanical problem $\dot{R}^{2}+V(R)=-k$, with $-k$
serving as the 'total mechanical energy'. 
We find
\begin{equation}
	-V(R)=\phi^{2}+\frac{\Lambda^{2}R^{4}}{81\phi^{2}}+
	\frac{2}{9}\Lambda R^{2} ~,
\end{equation}
where $54R\phi^{3}=\mu \pm \sqrt{\mu^{2}-4\Lambda^{3}R^{8}}$.
Depending on the size of $R$, our analysis bifurcates:

\noindent $\bullet$ If $4\Lambda^{3}R^{8}<\mu^{2}$, there exists
a single real solution
\begin{equation}
	-9V(R)=\left(\frac{\mu+\sqrt{}}{2R}\right)^{2/3}+
	\left(\frac{\mu-\sqrt{}}{2R}\right)^{2/3}+
	2\Lambda R^{2} ~,
	\label{VRsmall}
\end{equation}
where $\sqrt{}\equiv\sqrt{\mu^{2}-4\Lambda^{3}R^{8}}$, so we are
definitely on the $\xi>1$ branch (to be more accurate, $\xi>4/3$
in this case).

\noindent $\bullet$ If $4\Lambda^{3}R^{8}>\mu^{2}$, one has to be a
bit careful to pick up the solution which is the analytic continuation
of the above.
This $\xi>1$ solution (to be more accurate, $1<\xi<4/3$ in this
case) is given by
\begin{equation}
	-9V(R)=4\Lambda R^{2}\cos^{2}{\theta} ~,
	\label{VRbig}
\end{equation}
where ${\displaystyle \cos{3\theta}=\frac{\mu}{2\Lambda^{3/2}R^{4}}}$.

As advertised, we approach the universal behavior $
V(R)\simeq -\frac{1}{9}(\mu/R)^{2/3}$ at very short distances,
and recover the Einstein-de-Sitter parabola $V(R)\simeq -\frac{1}{3}
\Lambda R^{2}$ at very long distances.
$V(R)$ exhibits a maximum around
\begin{equation}
	R_{\Lambda}\approx \left(\frac{\mu^{2}}
	{27 \Lambda^{3}}\right)^{1/8} ~.
\end{equation}
This was the scale of the Universe when inflation took over.
One can furthermore estimate the time this happened, namely
$t_{\Lambda}\approx \frac{3\sqrt{3}}{4}\Lambda^{-1/2}$, which is notably
$\mu$-independent.

At this stage, our discussion bifurcates again.
The fate of the baby Universe depends on whether $k$ (the
curvature term cannot be neglected at this stage) is above
or below the critical (positive) value of
\begin{equation}
	k_{\Lambda}\equiv -V(R_{\Lambda})\approx
	\frac{4}{9}(3\mu^{2}\Lambda)^{1/4} ~.
	\label{k0}
\end{equation}

\noindent $\bullet$ If $k<k_{\Lambda}$, the whole trail of
evolution is classically permissible, and in some sense boring.
Up to the discussed RT modifications, which are significant at
the very early and the dilute stages, the evolution is almost
standard. 

\noindent $\bullet$ If $k>k_{\Lambda}$, on the other hand, the
situation is truly fascinating.
This case is characterized by a potential barrier which the
\textit{embryo} Universe must cross in order to survive.
Classically, this embryo does not have a chance, and is doomed
to collapse in a Big-Crunch.
Quantum mechanically, however, the situation highly resembles
$\alpha$-decay, and thus seems to nicely connect (as $\mu\rightarrow
0$) with Hawking's idea\cite{hawk} of quantum nucleation.
Hartle and Hawking, in their "no-boundary" proposal\cite{hh}, have
invoked the Euclidean region to altogether avoid the Big-Bang
singularity and the dependence on initial conditions.

\medskip
\noindent \textit{\textbf{Concluding remarks:}}

Is the cosmological 'missing' mass, also known as the
$\Omega$-puzzle, a signature of a brane-like Universe?
Attempting to answer this question in the affirmative, we
have traded Einstein gravity for Regge-Teitelboim gravity,
constituting an elegant and in principle testable deviation
from standard cosmology.
The emerging theory exhibits a built-in Einstein limit,
offers definite predictions (e.g. enhanced age of the Universe
and a reduced deceleration parameter), dictates a universal
Big Bang behavior, and is cosmological constant free.
The realistic $\Lambda \neq 0$ case is to be studied elsewhere.
We can report, in passing, the derivation\cite{DK} of a
quadratic Hamiltonian for string-like gravity, leading to a
bifurcated Wheeler-Dewitt-like equation.

\medskip
\noindent \textit{\textbf{acknowledgments:}}
Enlightening discussions with Professors P. Mannheim and E. Guendelman
are gratefully acknowledged.

\end{document}